# An example of regional seismicity recovery on June 22, 2020, using waveform cross-correlation

Ivan O. Kitov

**Abstract**

June 22nd, 2020 was an important day for nuclear test monitoring. The US undersecretary, Thomas DiNanno, made an official claim that a secret nuclear test was conducted at the Lop Nor testing site on that day. However, the Provisional Technical Secretariat (PTS) of the Comprehensive Nuclear-Test-Ban Treaty Organization (CTBTO) reported that there were no seismic events consistent with a nuclear explosion on that date. This means that the event was either hidden under the threshold of the international monitoring system (IMS), or it was missed by the international data center (IDC). Alternatively, no event occurred. All scenarios are of interest to the scientific community. The routine processing of seismic data by the IDC is well documented in open sources. The threshold for detection is not uniform across the globe, especially in continental areas. Additionally, it is not low enough to detect weapon-sized explosions using the cavity decoupling method. There are methods to lower this threshold using waveform cross-correlation (WCC) techniques. On June 22th, 2020, the IDC ran a preliminary pipeline for processing seismic data from the IMS. This was done in a testing mode. Since 2020, the development of WCC methods has allowed for revisiting cases and obtaining high-quality reports on events that were not reported five years ago.

## Introduction

For the nuclear tests monitoring community, the claim of the test conducted on June 22, 2020 is a professional challenge. The detection threshold worldwide is low enough to make sure that weapon-size test would be found with the probability high enough to prohibit an attempt to conduct such a hidden test. The Provisional Technical Secretariat of the Comprehensive nuclear-test-ban Treaty Organization (PTS CTBTO) includes the International Monitoring System (IMS) and International Data Centre to obtain and process data related to its mandate as per the Treaty [CTBT, 1996]. The broader monitoring community uses additional data, methods and technologies extending the capabilities of the PTS CTBTO. Among these methods, waveform cross-correlation (WCC) is an efficient way to reduce the detection threshold by an order of magnitude [Schaff, Richards., 2004; Bobrov et al., 2014; Kitov, Sanina, 2025].

The capability of the WCC-based detection and phase association as applied to the IMS data was proved in many experimental studies which used the IDC interactive review accomplished by IDC experienced analysts applied to the WCC automatic bulletin called XSEL - cross correlation standard event list. There were 50% to 70% reviewed events created extra to the Reviewed Event Bulletin (REB) of the IDC. These events matched the Event Definition Criteria of the IDC [Coyne et al., 2012] and had to be included in the REB, if such a procedure existed. Correspondingly, the current REB version does not include these newly created REB-ready events. The REB also does not include those REB-ready events which could be created according to the IDC rules using the XSEL out of the days and areas processed in the referred experimental studies. Therefore, the event on June 22, 2020 could be the one missed by the IDC and potentially found by the WCC method. Here, we reprocess IMS data from June 21 to June 23, 2020, using the most recent version of WCC-processing. The area is limited by coordinates 38°N-45°N, 84°E-93°E. This study is an attempt to demonstrate the method resolution and reliability and can be extended to any area of interest.

**Data and method**

The IMS seismic network includes primary and auxiliary stations sending data to the IDC in a continuous mode [Coyne et al., 2012]. There are array stations representing phased antennas class for seismic waves [Schweitzer et al., 2012] and three-component (3-C) stations. The IMS seismic network has global coverage and relies on high-frequency regional waves as the source of data to detect the events with the lowest magnitudes, such as the events conducted with in large underground cavities (this evasion method is also known as cavity decoupling) [Evernden et al., 1986]. In the final configuration consisting of 50 primary stations within continental regions, at least three of them have to be within 20° from any point at the surface. According to the CTBT, auxiliary seismic stations have to assist in the estimation of event parameters (i.e., epicenter, depth, magnitudes) but not to define the event statistical significance. Here, we use all available IMS stations to create event hypotheses and not to apply any screening criteria [CTBT, 1996].

Detection of primary P-phases at regional distances ($P_g$ and $P_n$) is the most efficient with the arrays having aperture of 1 to 10 kilometers, which are usually called small aperture arrays. However, the IMS has array stations with various apertures from several hundreds of meters (KVAR) to 40+ kilometers (NOA). The detection threshold of an array station is lower than that of a collocated 3-C station by a factor proportional to the square root of the number of the array elements. For a 16-element array, the gain is 4 times. In practice, the gain is lower due to beam loss and other physical phenomena [Schweitzer et al., 2014]. The availability of array stations allows the global monitoring system to be more sensitive (lower detection threshold) and have higher resolution (more accurate locations and magnitude estimations). The area of this study has several array stations within 30° degrees.

Standard IDC detector used for seismic waves is based on a robust approach related to the change in energy flux associated with signal arrival. To describe the energy flux the average amplitude of the ambient noise and potential signal are used. One can use the RMS amplitude or the absolute amplitude value to calculate the energy flux parameter in the long (long-term-average or LTA) and short (short-term-average or STA) windows. The STA/LTA ratio also used and signal-to-noise ratio (SNR) represents the difference between signal and noise, with the noise being a mixture of various signals of lower amplitude. The STA/LTA detector declares signal arrival when its value reaches the predefined threshold, which has to be estimated in an independent interactive procedure. This detector is a robust one in terms that it does not depend on the internal properties of the signal. But this feature also has its weakness as the signals differ in many other parameters. One of possibilities to improve the STA/LTA performance if waveform filtering which usually suppresses the noise signals in a larger proportion that the sought signal. Using a comb of filters, one often can find a filter efficient for a given detection case. The final list of detections at all IMS stations is then used to find the events generated them.

The phase association at the IDC consists in a global search of unique sources, as defined by their location, origin time and magnitudes together with the corresponding uncertainties, which likely generated the obtained arrivals [Coyne et al., 2012]. The event hypotheses created in a three-stage automatic Global Association (GA) procedure compile a Standard Event List (SEL3) serving as a start point for the interactive review conducted by IDC analysts. Approximately 60% of the events in the Reviewed Event Bulletin (REB), which is the final product of the interactive analysis, are based on the SEL3 seed events. Other REB events are obtained by different procedures including free searches from scratch. In the area 35°N-50°N, 80°E-100°E, the IDC has found more than 5000 events between the beginning of 2003 and the end of 2025. Figure 1 presents magnitude ($mb_{IDC}$) frequency distribution (0.2 magnitude unit bins). The catalog completeness threshold can be approximately estimated as $m_b$=4.0. The gap between the trend line below the threshold and the observation line increased with decreasing magnitude. This means that many real events are not found by the IDC standard methods. The previous experience with the WCC-based application demonstrate that a large portion of these missed events can be found in the IMS seismic data.

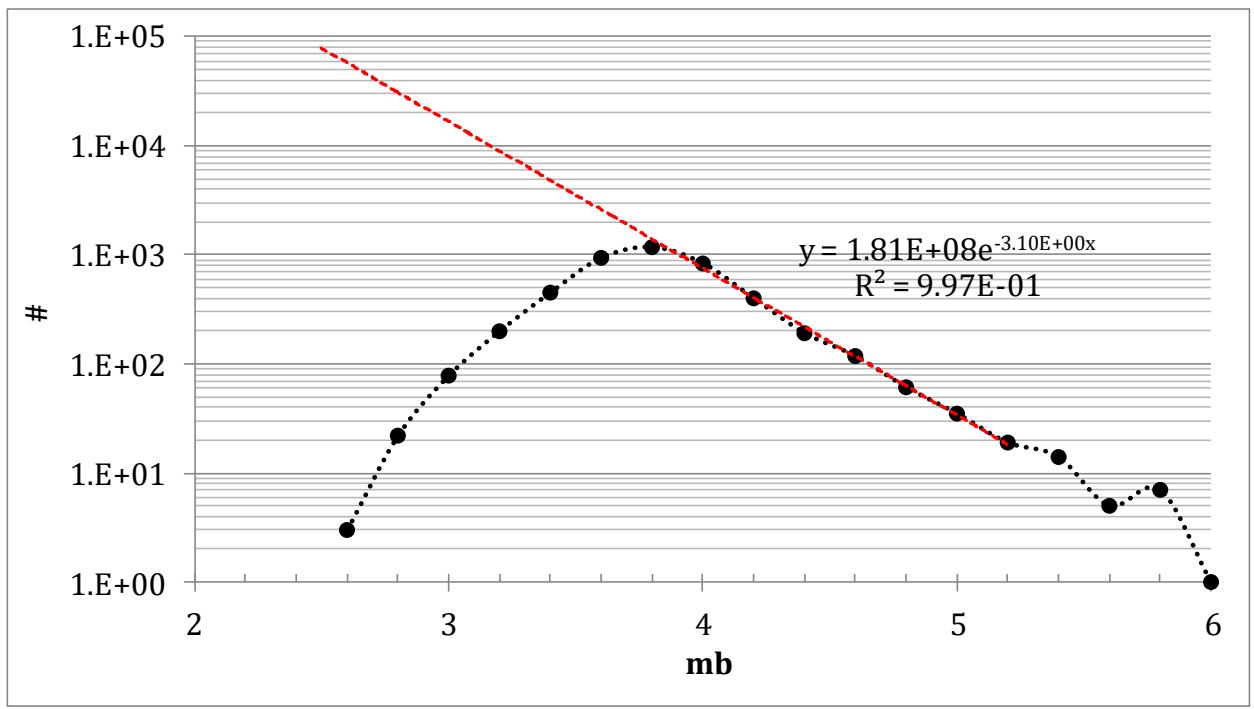

Figure 1. Magnitude frequency distribution as per REB.

The WCC-based detector uses the similarity of a sought signal with a waveform template obtained from an REB (master) event in the search area. It is implied by the procedure that the sought signal and the template are generated by events close in space to reach the needed level of similarity. The allowed distance between the master and sought event depends on the distance from both to the station and the apparent wave velocity. For regional event-station ranges, the master-sought event distance can be several kilometers, and for far-teleseismic ranges - tens of kilometers are still allowed. Array stations have an advantage of the correlation dependence on the azimuth difference between the sought and template signals as the arrival time difference at the array's individual elements can effectively destroy correlation.

However, the similarity between template and sought event shapes is not the only defining parameter. In the matched filter approach [Turin, 1960], the noise has to be stochastic and additive to maximize the SNR. The real ambient noise is far from the stochastic realization and the WCC detector performance critically depends on the noise properties. Overall, the correlation between template and sought signals has to be better than their correlation with the noise around the sought signal. In practice, a signal embedded in a computer generated quasi-stochastic noise can be found by itself (auto-correlation) even when its amplitude is ~1/10 of the noise. At the same time, the signals from aftershocks of biggest earthquakes are very difficult to detect as the ambient noise consist of similar and thus highly coherent signals. The studied area is characterized by relatively high seismicity and detection of weak signals is not in the optimal matched filter conditions.

The phase association procedure for the WCC detections is local in the sense that the sought event has to be close to the master. The empirical travel time from the sought event to the station can be calculated very accurately as a correction for the relative master-sought event positions and the slowness of the corresponding phase detected at the station. The local association (LA) procedure also has an advantage to use the probabilities of detections at the associated stations. The REB events can be used to calculate the frequency of station association with events in a small area around the master event (say, ~3°). The obtained probabilities can be considered as station weights, When the LA is applied to the arrivals detected by a single master, a requirement of minimum sum of the associated station weights, or event weight, is the threshold of the statistical significance of the event hypothesis. The events cannot consist of arrivals at stations with poor performance. The thresholds of event weight depend on the REB statistics and can be calculated for every REB event or for a global grid with the predefined node spacing. The latter approach smoothes the LA threshold distribution and allows to avoid problems with event wrong location and phase association.

In this study, the WCC detector and LA procedure are tuned to the area in a preliminary study. The allowed detection rate of 1 per 60 second defined the detection threshold at every station. The LA creates event hypotheses with the allowed travel time residuals of 2 s, as adopted by the IDC for the REB events. When the neighboring master event create hypotheses competing for the same physical arrival, the winner is defined by larger hypothesis weight. When the weights are equal, the number of associated phases is the decider, and the last resort is the RMS travel time residual.

**Results**

The day 22 June, 2020 is mentioned in the news as the event day. Since this date came from the news, it is not clear what time was used - local to the source, local to the news source, UTC. To mitigate the related uncertainty, three days were processed with the WCC - from June 21 to June 23, 2020. The master events were selected from the REB by a procedure based on a global grid of nodes with 1.5° spacing. The masters have to have the largest event weight among all other events within 0.75° of the grid node. These are the best events by weight. A more reliable procedure used to create a master event set for the routine WCC processing at the IDC was based on mutual cross-correlations between all events within the same circle and the winning master was the one which has found the largest portion of the other events under consideration for the given node. To improve the best master selection, a second search run was conducted with the first run winners excluded. The final set included 76 events as shown in Figure 2, which also illustrates seismicity within the studied area and adjacent regions as per REB. There are more than 5000 events recorded over a 20-year interval.

The XSEL bulletin obtained is presented in Figure 2. It includes 30 XSEL events: 12 on June 21, 12 on June 22, and 6 on June 23. It is worth mentioning that there was only one event between 0 and 6 a.m. (5:55 a.m. on June 21) , three events between 18 and 23 p.m. The period between 6 a.m. and 15 p.m. included 20 events. The timing of the events hints that the majority of detected events is related to open pit mining and other industrial activities. The XSEL events are created mainly in the spots already known by their historical activity. There are two active spots near the north-western and south-eastern corners of the studied area well defined by the shown nodes of the global grid with active master events.

The only REB event occurred at 10:38:07.04 (UTC), on June 22. It had 19 associated P-phases, $m_b$=3.87, depth 0 km, and epicenter 43.803°N, 86.773°E. This event is shown in Figure 2 together with the master event which found it and the XSEL solution, which is relatively far away from the REB. One likely need a better master event to get more accurate XSEL solution for the REB.

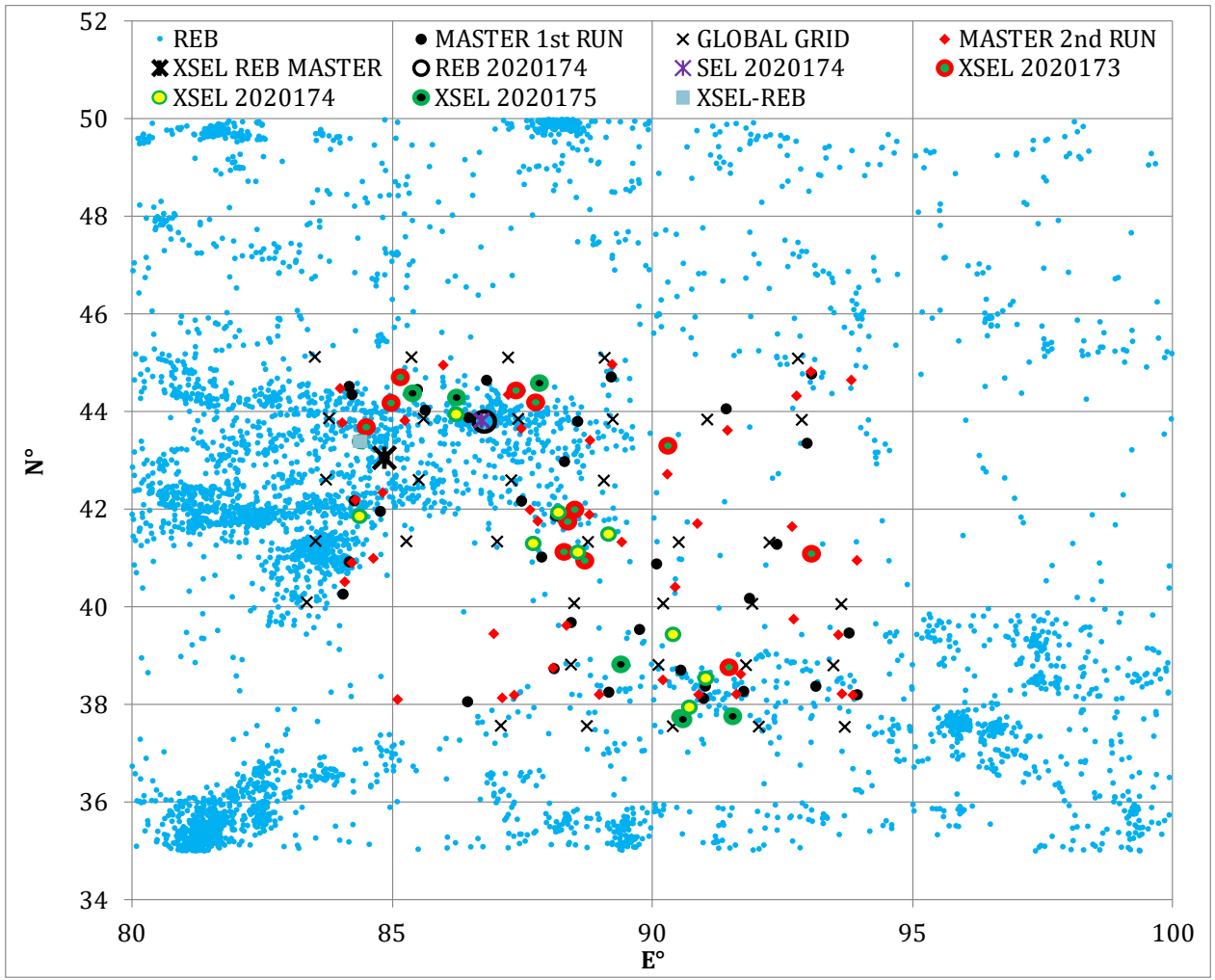

Figure 2. Seismicity of the region as per REB between 2003 and 2026. Master events obtained in two runs. Thirty XSEL events. The REB event on June 22 together with the master created the matching XSEL event.

**Discussion**

The REB event detected on June 22, 2020 is too strong to be misinterpreted as a potential test. There were no other events in the REB in the studied area for one day before and 2 days after June 22. The XSEL events are found within the magnitude range where the REB in incomplete and in the spots with higher seismic activity, natural or industrial. The XSEL events are created automatically and this have to pass interactive analysis according to the IDC before their REB readiness is confirmed. At the same time, the seismicity of the area is definitely higher at the magnitude levels below the IDC detection thresholds.

One of possible ways to reduce the seismic energy generation by an explosive source is the use of large underground cavity. This method is called cavity decoupling and was tested by the USA [Springer at al., 1968] and Soviet Union [Adushkin et al., 1993 ]. The cavity has to be spherical to provide the largest decrease in seismic efficiency, but the conduction of chemical explosion in elongated cavities similar to tunnels also provides a significant decoupling factor of around 5 to 10 [Murphy et al., 1997]. The cavity decoupling was a successful method and allowed to hide the Azgir-3-2 event conducted in the Azgir-3 cavity from seismic observations external to the USSR. Therefore, there exists an opportunity to hide an event with ~1kt yield in a cavity of approximately 20 m in radius. However, the evaporated rock may always seep to the surface through cracks in hard rocks and despite seismic invisibility the event will be easily detected. The full isolation of the debris can be reached when an event is detonated in the salt deposit with thick impermeable layers. The areas with such deposits would be of interest to process using the WCC-based methods.